\documentclass[twocolumn,showpacs,preprintnumbers,amsmath,amssymb]{revtex4}

\usepackage{graphicx}

\begin{document}
\title{\bf Entanglement of a Single Spin-1 Object: \\ An Example of Ubiquitous Entanglement}

\author{Sinem Binicio\v{g}lu$^1$, M. Ali Can$^2$, Alexander A. Klyachko$^3$, and Alexander S. Shumovsky$^1$}

\affiliation{$^1$Department of Physics, Bilkent University,
Bilkent, Ankara, 06800 Turkey \\ $^2$Jet Propulsion Laboratory,
California Institute of Technology, Pasadena CA 91109-8099, USA
\\ $^3$Department of Mathematics, Bilkent University, Bilkent, Ankara, 06800 Turkey}

\begin{abstract}
Using a single spin-1 object as an example, we discuss a recent
approach to quantum entanglement \cite{Kl-Sh-2006}. The key idea
of the approach consists in presetting of basic observables in the
very definition of quantum system. Specification of basic
observables defines the dynamic symmetry of the system. Entangled
states of the system are then interpreted as states with maximal
amount of uncertainty of all basic observables. The approach gives
purely physical picture of entanglement. In particular, it
separates principle physical properties of entanglement from
inessential. Within the model example under consideration, we show
relativity of entanglement with respect to dynamic symmetry and
argue existence of single-particle entanglement. A number of
physical examples are considered.
\end{abstract}


\maketitle

\section{Introduction}

Recent development of quantum information technologies has led to
a number of successful and promising realizations of protocols
based on the use of quantum entanglement. For example, the quantum
key distribution has recently become an industrial product
\cite{QKD}. These developments caused a great burst of activity in
investigation of quantum entanglement. During the last decade, the
applications of entanglement to quantum information and quantum
computing were discussed in a huge number of articles and review
papers, in particular in \emph{Foundations of Physics}
\cite{foun}.

Nevertheless, the substance of entanglement still remains unclear,
especially beyond the simplest case of two-qubit systems.
Moreover, there is a certain muddle even in the very definition of
entanglement because important and inessential are often jumbled
together.

In this paper, our aim is not to discuss the applications but the
physics behind the quantum entanglement.

For example, entanglement is usually associated with quantum {\it
nonlocality} or violation of classical realism
\cite{EPR,Bell,Nielsen}. This simply means that measurements on
{\it spatially separated} parts of a quantum system may
instantaneously influence one another. Physically this is caused
by the quantum correlations between the parts of the
system\cite{Bell}. Once created, those correlations keep on
existing even after the spatial separation of parts.

On one hand, the nonlocality is probably the main distinguishing
feature of quantum mechanics regarding classical physics. On the
other hand, this notion does not contain any quantification of
distance between separated entangled parts of a quantum system.
Thus, it seems to be natural to assume that quantum system with
strongly correlated intrinsic parts may manifest entanglement
independent of distance between the parts and hence even as a
local object without spatial separation of parts
\cite{Kl-Sh-2003,Enk,Viola-2004,Can-2005,Kl-Sh-2006}.

The quantum nonlocality is often expressed in terms of violation
of different Bell-type conditions of classical realism
\cite{Bell}. This violation is a characteristic feature of
entanglement in two-qubit systems. However, unentangled states of
some systems beyond two qubits can also manifest the violation of
those conditions \cite{Klyachko-2002,Barnum,Klyachko-2006}. For
example, the difference between entangled and unentangled states
disappears for systems with dynamic symmetry group
$\mathrm{SU}(\mathcal{H})$ with dimension of the Hilbert space
$\dim \mathcal{H}\geq 3$ (see Ref. \cite{Klyachko-2007}, cf.
\cite{Peres}). As a matter of fact, violation of Bell-type
conditions generally indicates the absence of ``hidden" classical
variables in quantum mechanics \cite{Bell} rather than
entanglement (also see Appendix A).

This allows us to conclude that nonlocality and violation of
classical realism alone are not the essential sign of entanglement
and that there is no physical prohibition for the existence of
entanglement of local objects (particles) caused by quantum
correlations of their {\it intrinsic degrees of
freedom}\cite{Kl-Sh-2003,Enk,Viola-2004,Can-2005,Kl-Sh-2006}.

Another common opinion is that the entanglement of multipartite
systems defined in the Hilbert space $\mathcal{H}=\mathcal{H}_A
\otimes \mathcal{H}_B \otimes \cdots$ can be associated with the
{\it nonseparability} of states $\psi \in \mathcal{H}$ with
respect to the parts of the system (e.g., see \cite{Bruss}).

This statement, which is undoubtedly valid in the case of
bipartite systems, does not have a lucid sense for multipartite
entanglement. Stress that the notion of nonseparability is clearly
useless in the case of single-particle entanglement,
\cite{Viola-2004,Can-2005}.

Three-qubit states, whose classification has been constructed in
Ref. \cite{Miyake} can be considered as an example. Namely, the
states from the class, specified by the nonseparable GHZ
(Greenberger-Horne-Zeilinger) state, manifest only three-partite
entanglement with no correlations (entanglement) between pairs of
qubits. In contrast, the nonseparable W-states manifest
entanglement only between pairs of qubits while they are
unentangled in the three-qubit sector. In turn, the so-called
bi-separable states (separable with respect to one of the parties)
may manifest bipartite entanglement. For details, see Appendix B.

Thus, it seems reasonable to set aside the criteria of
entanglement based on nonlocality, violation of classical realism,
and nonseparability, and to focus attention on physical
manifestations of entanglement in the process of measurement of
quantum observables.

Note first that there is a certain interdependence between quantum
correlations peculiar to entangled states and quantum
uncertainties (fluctuations) of {\it local} observables
\cite{Klyachko-2002,Can-2002,Fluctuations}. Consider as an
illustrative example the measurement of spin projection onto the
quantization axis in the two-qubit states
$|\psi_{\uparrow\uparrow} \rangle =|\uparrow\uparrow \rangle$ and
$|\psi_{CE} \rangle =(|\uparrow\uparrow \rangle
+|\downarrow\downarrow \rangle)/\sqrt{2}$. For the correlation
functions and variances (uncertainties), we get
\begin{eqnarray}
\begin{array}{ll}
\langle \psi_{\uparrow\uparrow}| \sigma_z^A; \sigma_z^B|
\psi_{\uparrow\uparrow} \rangle =0, &
V(\sigma_z^{A,B};\psi_{\uparrow\uparrow})=0, \\ \langle \psi_{CE}|
\sigma_z^A;\sigma_z^B|\psi_{CE} \rangle =1, &
V(\sigma_z^{A,B};\psi_{CE})=1. \end{array} \nonumber
\end{eqnarray}
Here $\sigma_z^A$, $\sigma_z^B$ denote the $z$-component of Pauli
spin operator,
\begin{eqnarray}
\langle \psi|\sigma^A;\sigma^B|\psi \rangle = \langle
\psi|\sigma^A \sigma^B|\psi \rangle - \langle \psi|\sigma^A|\psi
\rangle \langle\psi|\sigma^B|\psi \rangle \nonumber
\end{eqnarray}
is the correlation function of local measurements, and
\begin{eqnarray}
V(\sigma;\psi)= \langle \psi|\sigma^2|\psi \rangle - \langle
\psi|\sigma|\psi \rangle^2 \nonumber
\end{eqnarray}
is the variance of the observable $\sigma$ in the state $\psi$.
Thus, the correlation functions and variances have similar
behavior for the characteristic states like
$\psi_{\uparrow\uparrow}$ and $\psi_{CE}$.

The natural question now is how many physical observables should
be measured in order to conclude that a given state of a certain
system is entangled \cite{Huelga}? This question has extremely
high importance for understanding of physical essence of
entanglement and its quantification. Besides that, this question
has a quite practical meaning in connection with test of sources
of entangled states \cite{Hayashi}.

In a recent approach \cite{Klyachko-2002,Kl-Sh-2003,Kl-Sh-2004}
(for recent review, see Ref. \cite{Kl-Sh-2006}), it has been
proposed to begin the analysis of entanglement with the choice of
independent {\it basic observables} that can be associated with
the orthogonal basis of a certain Lie algebra $\mathcal{L}$. The
corresponding Lie group $G=\exp(i\mathcal{L})$ defines the {\it
dynamic symmetry} of the physical system under consideration.

It should be emphasized that the idea to specify a quantum system
by accessible observables is known for a long time (e.g., see
\cite{Hermann}). Unfortunately, this principle idea is often set
aside. As we show below in this paper, this principle plays
extremely important role in description of quantum entanglement.

Within the approach of Refs.
\cite{Klyachko-2002,Kl-Sh-2003,Kl-Sh-2004}, the {\it complete
entanglement} is interpreted as {\it manifestation of quantum
uncertainties of all basic observables at their extreme}. By
complete entanglement we mean here the maximal entanglement that
can be achieved by pure states.

Note that, for a given quantum system, it is enough to know the
completely entangled states because all other entangled states can
be generated from those states through the use of SLOCC
(stochastic local operations assisted by classical communications)
\cite{Dur,Verstraete}.

The dual objective of present paper is to discuss the
characteristic features of this approach, using a single {\it
qutrit} (ternary quantum state) as an illustrative example of some
considerable interest, and hence to justify existence of
single-particle entanglement.

Qutrit is usually associated with ternary unit of quantum
information \cite{Klyshko}. Instructiveness of this example
consists in the {\it relativity of entanglement} with respect to
the choice of dynamic symmetry $\mathrm{G}$ of ternary quantum
physical system.  Namely, one can choose either
$\mathrm{G}=\mathrm{SU}(3)$ \cite{Caves} or
$\mathrm{G}'=\mathrm{SU}(2)$. Just the latter case of a single
spin-1 system may manifest entanglement without division of the
system into separated parts \cite{Kl-Sh-2003,Viola-2004,Can-2005}.

Stress that entanglement of a single photon has been discussed for
a long time \cite{single-photon}. The picture always involves an
external qubit formed by two possible paths owing to its
propagation through a beam splitter. In contrast, our concept of
the single-particle entanglement \cite{Can-2005,Kl-Sh-2006}
considers particle itself independent of its environment (see also
important discussions in Refs. \cite{Viola-2004,Enk}). In this
case, quantum correlations peculiar to entanglement can be
associated with {\it intrinsic degrees of freedom} of a particle.

In the paper, we show that the single-particle entanglement has
all important properties of conventional two-qubit entanglement.
In particular, its unentangled states are the spin-coherent states
like those of two qubits \cite{Klyachko-2002,Barnum} (concerning
spin-coherent states, see the basic works \cite{Coherence-A} and
monographs \cite{Coherence-B}). In turn, the single-particle
entangled states are squeezed like the two-qubit entangled states
(for relation between squeezing and ``conventional" multipartite
entanglement, see Refs. \cite{Squeezing}).

Note that association of unentangled and coherent states on the
one hand and entangled and squeezed states on the other hand
blends well with the concept of entanglement as manifestation of
quantum uncertainties at their extreme.

We also reveal the mechanism of intrinsic quantum correlations
hidden behind the single-qutrit entanglement.

The paper is organized as follows. In Section II, the principles
of the dynamic symmetry approach to quantum entanglement are
presented and we show existence of the single-particle
entanglement for spin-1 systems. In Section III, we discuss
properties of that entanglement. Then, in Section IV, we consider
some physical realizations. The conclusions are given in Section
V.

\section{Entanglement and quantum fluctuations}

As we have said in Introduction, specifying a given quantum
system, we should first choose the accessible independent physical
observables associated with dynamic symmetry of the system
\cite{Klyachko-2002,Kl-Sh-2006}.

For example, in the case of a qubit (spin 1/2) system, dynamic
symmetry is given by the group $\mathrm{SU}(2)$. The orthogonal
basis of the corresponding Lie algebra $su(2)$ consists of three
spin operators (Pauli matrices). Thus, a two-qubit system is
characterized by the dynamic symmetry $\mathrm{G}=\mathrm{SU}(2)
\times \mathrm{SU}(2)$, which corresponds to the six basic
observables (three Pauli matrices for each part). For the
two-qubit pure state, the number of necessary measurements,
providing information about entanglement carried by this state, is
reduced to three \cite{Huelga} because of the local character of
the measure of entanglement (concurrence) in this case
\cite{Buzek}.

To illustrate special importance of the specification of quantum
system by basic observables, consider a qutrit (ternary unit of
quantum information) associated with a state
\begin{eqnarray}
|\psi \rangle = \sum_{s =-1}^1 \psi_{s} |s \rangle, \quad
\sum_{s=-1}^1 |\psi_{s}|^2=1 \label{qutrit}
\end{eqnarray}
in the three-dimensional Hilbert space $\mathcal{H}_3$. As we have
mentioned in Introduction, there are at least two qualitatively
different physical systems, whose states are qutrits. Namely, one
possible realization corresponds to the general symmetry
$\mathrm{G}=\mathrm{SU}(3)$ of the system, which implies eight
basic observables (Gell-Mann matrices) \cite{Caves}

\begin{widetext}
\begin{eqnarray}
\lambda_1= \left( \begin{array}{ccc} 0 & 1 & 0 \\ 1 & 0 &  0 \\ 0
& 0 & 0
\end{array} \right), \quad \lambda_2= \left( \begin{array}{crr} 0 & -i & 0 \\ i &
0 & 0 \\ 0 & 0 & 0 \end{array} \right) , \quad \lambda_3= \left(
\begin{array}{crr} 1 & 0 & 0 \\ 0 & -1 & 0 \\ 0 & 0 & 0
\end{array} \right) , \quad \lambda_4= \left( \begin{array}{ccc} 0 & 0 & 1 \\ 0 & 0 & 0 \\ 1 &
0 & 0 \end{array} \right) , \nonumber \\
 \lambda_5= \left(
\begin{array}{ccr} 0 & 0 & -i
\\ 0 & 0 & 0 \\ i & 0 & 0 \end{array} \right) , \quad \lambda_6=
\left( \begin{array}{ccc} 0 & 0 & 0 \\ 0 & 0 & 1 \\ 0 & 1 & 0
\end{array} \right) ,  \quad \lambda_7= \left( \begin{array}{ccr} 0 & 0 & 0 \\ 0 &
0 & -i \\ 0 & i & 0 \end{array} \right) , \quad \lambda_8=
\frac{1}{\sqrt{3}} \left( \begin{array}{ccr} 1 & 0 & 0 \\ 0 & 1 &
0 \\ 0 & 0 & -2
\end{array} \right) . \label{Gell-Mann}
\end{eqnarray}
\end{widetext} Hereafter we call the corresponding system the {\it
true qutrit} system.

Another realization assumes reduced symmetry
$\mathrm{G}'=\mathrm{SU}(2)$ of the physical system, which
requires only three basic observables (spin-1 operators)
\cite{Can-2005}

\begin{widetext}
\begin{eqnarray}
S_x =  \frac{1}{\sqrt{2}} \left( \begin{array}{ccc} 0 & 1 & 0 \\ 1 & 0 & 1 \\
0 & 1 & 0 \end{array} \right) , \quad S_y =  \frac{1}{\sqrt{2}}
\left( \begin{array}{rrr} 0 & -i & 0
\\ i & 0 & -i \\ 0 & i & 0
\end{array} \right) , \quad S_z =  \left( \begin{array}{ccr} 1 & 0 &
0 \\ 0 & 0 & 0 \\ 0 & 0 & -1 \end{array} \right) . \label{spin-1}
\end{eqnarray}
\end{widetext}

We call this case the {\it spin-qutrit} system.

As we discuss in the next Section, qutrit (\ref{qutrit}) may
manifest entanglement in the case of single spin-qutrit system,
while single true qutrit can never be entangled.

Before we begin to consider entanglement of a single qutrit, let
us briefly discuss the physical definition of entanglement of
Refs. \cite{Klyachko-2002,Kl-Sh-2003,Kl-Sh-2004}.

For a given state $\psi$ of a system with basic observables $X_i$,
we can measure the expectation values $\langle \psi|X_i|\psi
\rangle$ and variances (uncertainties)

\begin{eqnarray}
V(X_i;\psi)= \langle \psi|X_i^2|\psi \rangle -\langle
\psi|X_i|\psi \rangle^2 . \label{variance}
\end{eqnarray}

It is interesting that Wigner and Yanase \cite{Wigner-Yanase} have
proposed a new quantity to measure specific quantum information
about a state $\psi$, that can be obtained from {\it macroscopic
measurement} of the observable $X_i$ in this state (the so called
{\it Wigner-Yanase skew information}):

\begin{eqnarray}
I_{WY}(X_i;\rho)=-\frac{1}{2}Tr([\sqrt{\rho},X_i]^2) . \label{ske}
\end{eqnarray}
In the case of pure state $\psi$, the density matrix takes the
form $\rho=|\psi\rangle\langle\psi|$, the quantity (\ref{ske})
simply coincides with the variance (\ref{variance}) (see Refs.
\cite{Wigner} for further discussion of Wigner-Yanase quantum
``skew information").

Following \cite{Klyachko-2002,Kl-Sh-2003,Kl-Sh-2004}, introduce
the {\it total variance}
\begin{eqnarray}
\mathbb{V}(\psi)= \sum_i V(X_i;\psi) \label{total-variance}
\end{eqnarray}
calculated for all basic observables and all parts of the system
(in the case of multipartite systems). By definition, this
quantity (\ref{total-variance}) is an invariant, independent of
the choice of basis of the Lie algebra $\mathcal{L}$ of
observables. This quantity (\ref{total-variance}) can also be
interpreted as the total amount of Wigner-Yanase information
peculiar to the state $\psi$.

It was proposed in Refs.
\cite{Klyachko-2002,Kl-Sh-2003,Kl-Sh-2004} that, {\it complete
entangled states} $\psi_{CE}$ of an arbitrary system can be
defined in terms of maximum of total variance:
\begin{eqnarray}
\mathbb{V}(\psi_{CE})= \max_{\psi \in \mathcal{H}}
\mathbb{V}(\psi). \label{complete-entanglement}
\end{eqnarray}
This definition has a simple physical meaning. It associates
complete entanglement with the maximal amount of quantum
uncertainty in a given system. Validity of this definition in some
known cases of completely entangled states of multipartite systems
has been shown in a number of papers (see Ref. \cite{Kl-Sh-2006}
for references).

It is seen that Eq. (\ref{complete-entanglement}) represents a
certain variational principle, similar in a sense to the maximal
entropy principle in statistical physics, which is used to define
equilibrium states.

At first glance, Eq. (\ref{complete-entanglement}) defines only
completely entangled states $\psi_{CE}$. In fact, it can be used
to specify all entangled pure states of the system as well. The
point is that all entangled states of a given system are
equivalent to SLOCC (stochastic local operations assisted by
classical communications) \cite{Dur,Verstraete}. Note that SLOCC
are represented by operators from the complexified dynamic
symmetry group \cite{Verstraete}
\begin{eqnarray}
\widehat{SLOCC} \equiv g^c \in G^c= \exp(\mathcal{L} \otimes
\mathbb{C}). \nonumber
\end{eqnarray}
Thus, for the entangled states $\psi_E$ we get
\begin{eqnarray}
|\psi_E \rangle =g^c|\Psi_{CE}\rangle. \label{entangled-state}
\end{eqnarray}

Note that in the case of compact Lie algebra (like
$\mathrm{SU}(N)$), the quadratic form
\begin{eqnarray}
\sum_i X_i^2=C_{\mathcal{H}} \nonumber
\end{eqnarray}
is a scalar (Casimir operator). Then Eq. (\ref{total-variance})
takes the form
\begin{eqnarray}
\mathbb{V}(\psi)=C_{\mathcal{H}}- \sum_i \langle \psi|X_i|\psi
\rangle^2. \label{total-variance-a}
\end{eqnarray}
It is easily seen that the maximum of the total variance
(\ref{total-variance-a}) is provided by the condition
\begin{eqnarray}
\forall i \quad \langle \psi_{CE}|X_i|\psi_{CE} \rangle =0.
\label{condition-CE}
\end{eqnarray}
This condition represents a set of algebraic equations for the
complex coefficients of the wave function $|\psi\rangle$, which
enables us to fairly simplify the analysis of entanglement.
Validity of this condition  (\ref{condition-CE}) for completely
entangled qubit-states in quite general settings has been checked
in Ref. \cite{Can-2002}. Because the condition
(\ref{condition-CE}) deals directly with measurement of physical
observables, it has been proposed in Ref. \cite{Can-2002} to use
the condition as an {\it operational definition} of complete
entanglement.

Amount of entanglement carried by entangled states
(\ref{entangled-state}) can also be measured by means of total
variance as follows \cite{Baris}
\begin{eqnarray}
\mu(\psi)=
\sqrt{\frac{\mathbb{V}(\psi)-\mathbb{V}_{\min}}{\mathbb{V}_{\max}-\mathbb{V}_{\min}}}
. \label{measure}
\end{eqnarray}
Here $\mathbb{V}_{\max}$ and $\mathbb{V}_{\min}$ denote the total
variance for completely entangled and unentangled states,
respectively. This measure coincides with the {\it concurrence}
\cite{Concurrence} for pure states of an arbitrary bipartite
system. It can also be applied beyond bipartite systems. For
unentangled states, $\mu(\psi)=0$, while for entangled states it
lies in $(0,1]$, so that $\mu (\psi_{CE})=1$.

\section{Entanglement in a single spin-qutrit system}

Note that the definition of complete entanglement
(\ref{complete-entanglement}) and its equivalent form
(\ref{condition-CE}) do not assume the multipartite character of
quantum systems.

Does the single qubit obey the condition (\ref{condition-CE})? The
answer is not. The point is that the pure single-qubit state
\begin{eqnarray}
\psi =a|\uparrow \rangle +b|\downarrow \rangle ,
\quad|a|^2+|b|^2=1 \nonumber
\end{eqnarray}
is in fact characterized by only two real parameters ($|a|$ and
$\arg a- \arg b$), for which three Eqs. (\ref{condition-CE}) with
Pauli matrices as basic observables have only trivial solution.

For decades, qubits remain the main object of quantum information.
Therefore, nonexistence of single-qubit entanglement is frequently
used as a general argument against the single-particle
entanglement (see nice discussion in Ref. \cite{Enk}).

We now turn to the qutrit (\ref{qutrit}), which is specified by
five real parameters. Equations (\ref{condition-CE}) with eight
basic observables (\ref{Gell-Mann}) clearly have only trivial
solutions, so that single true qutrit system does not manifest
entanglement like single qubit system.

Situation changes qualitatively if qutrit (\ref{qutrit}) is
considered as a state of spin-qutrit system with only three basic
observables (\ref{spin-1}) \cite{Can-2005}. In this case,
equations (\ref{condition-CE}) with three spin-1 operators
(\ref{spin-1}) have nontrivial solutions, so that complete
entanglement of a single spin qutrit system is allowed. In
particular, it is straightforward to calculate the measure
(\ref{measure}) for the single spin-qutrit state:
\begin{eqnarray}
\mu(\psi)=2|\psi_{-1} \psi_1-\psi_0^2/2|. \label{measure-qutrit}
\end{eqnarray}
Thus, the state (\ref{qutrit}) of a single spin-1 system manifests
entanglement if its coefficients obey the condition
\begin{widetext}
\begin{eqnarray}
\frac{1}{4} \geq |\psi_{-1}|^2|\psi_1|^2+\frac{1}{4} |\psi_0|^2-
|\psi_{-1}||\psi_0||\psi_1|^2 \cos(\phi_{-1}+\phi_{1}-2\phi_0)
>0. \label{qutrit-entanglement}
\end{eqnarray}
\end{widetext}
Here $\phi_{\ell}=\arg \psi_{\ell}$. Complete entanglement is
achieved when this form (\ref{qutrit-entanglement}) takes the
value $1/4$. For example, the states
\begin{eqnarray}
|\psi_0 \rangle = |0 \rangle \label{s=0-state}
\end{eqnarray}
and
\begin{eqnarray}
|\psi_{\pm} \rangle = \frac{1}{\sqrt{2}} (|1 \rangle \pm |-1
\rangle ) \label{s=1-states}
\end{eqnarray}
are completely entangled qutrit states of a single spin-qutrit
system.

Before we begin to discuss the precise meaning of the above
obtained result, let us stress the {\it relativity of
entanglement} with respect to dynamic symmetry of physical system.
The same state (\ref{qutrit}) is unentangled if dynamic symmetry
of the system is $\mathrm{G}=\mathrm{SU}(3)$ and entangled in the
case of reduced dynamic symmetry $\mathrm{G}'=\mathrm{SU}(2)$.

To interpret entanglement of single spin-qutrit system, let us
compare it with two-qubit entanglement that has been scrutinized
thoroughly.

It is known that an entangled two-qubit state is associated with
the $\mathrm{SU}(2)$ squeezed states \cite{Squeezing}, while
unentangled states are the $\mathrm{SU}(2)$ coherent states
\cite{Klyachko-2002,Barnum}. We now show that this interpretation
is valid for the entangled and unentangled states of a single
spin-qutrit as well.

Let us begin with the $\mathrm{SU}(2)$ coherent states that, for a
spin $s$, are defined in the following way
\cite{Coherence-A,Coherence-B}
\begin{eqnarray}
|\alpha \rangle =D(\alpha)|-s \rangle, \quad \alpha \in
\mathbb{C}, \label{coherent-state}
\end{eqnarray}
where
\begin{eqnarray}
D_\alpha = \exp (\alpha S_+- \alpha^* S_-) \label{displacement}
\end{eqnarray}
and $|-s \rangle$ is the lowest state among the $(2s+1)$ states of
spin-$s$ system. Here
\begin{eqnarray}
S_{\pm}=S_x \pm iS_y \nonumber
\end{eqnarray}
are the spin rising and lowering operators, respectively.

In the ``vacuum" state $|-s \rangle$, the spin has a given
projection $-s$ onto the $z$-axis $\langle -s|S_z|-s \rangle =-s$,
so that the corresponding variance $V(S_z;-s)=0$. For the two
other spin operators in the direction orthogonal to the
quantization axis $z$ we get
\begin{eqnarray}
\langle -s|S_{x}|-s \rangle= \langle -s|S_{y}|-s \rangle = 0,
\nonumber \\ V(S_{x};-s)=V(S_{y};-s)=s/2, \nonumber
\end{eqnarray}
so that the total variance (\ref{total-variance}) takes the from
\begin{eqnarray}
\mathbb{V}(-s)=s. \nonumber
\end{eqnarray}
This is the minimal value of the total variance for the spin-$s$
system under consideration. Thus, in view of the definition of
entanglement, given in the previous Section, the state $|-s
\rangle$ is unentangled.

According to Eq. (\ref{total-variance-a}), the maximum of the
total variance of a single spin-$s$ system is
\begin{eqnarray}
\mathbb{V}_{max}=\mathbb{V}(\psi_{CE})=s(s+1). \nonumber
\end{eqnarray}
This allows us to represent the measure of entanglement
(\ref{measure}) for a single spin-$s$ system in the following form
\begin{eqnarray}
\mu(\psi)= \frac{1}{s} \sqrt{\mathbb{V}(\psi)-s}.
\label{measure-spin-s}
\end{eqnarray}
Thus, the measure (\ref{measure}) vanishes for coherent states.

It is easily seen that, in the case of a single qubit ($s=1/2$),
any state of the system is a coherent one. While in the case of
single spin-qutrit ($s=1$), coherent states (\ref{coherent-state})
have the form
\begin{widetext}
\begin{eqnarray}
|\alpha \rangle =\frac{e^{2i \phi}}{2} [1-\cos
(2|\alpha|)]|+1\rangle +\frac{e^{i \phi}}{\sqrt{2}}
\sin(2|\alpha|)|0 \rangle +  \frac{1}{2} [1+\cos(2|\alpha|)]|-1
\rangle , \label{spin-1-coherent-state}
\end{eqnarray}
\end{widetext}
where $\phi = \arg \alpha$. Substituting
coefficients of the state (\ref{spin-1-coherent-state}) into Eq.
(\ref{measure-qutrit}), we can see that the measure of
entanglement vanishes like in the case of state $|-1 \rangle$.
This is natural. The point is that the operator $(\alpha
S_+-\alpha^* S_-)$ in (\ref{displacement}) belongs to the $su(2)$
algebra, so that the displacement operator (\ref{displacement})
amounts to an $\mathrm{SU}(2)$ rotation. This means that every
spin coherent state (\ref{coherent-state}) is just a state with
minimal spin projection $-s$ onto some direction, which can be
chosen as a new quantization axis. Thus, there is no principle
difference between the spin coherent state and state $|-s\rangle$.
In particular, spin coherent state is as unentangled as the state
$|-s\rangle$.

Consider now the spin squeezed state for the spin-qutrit system
under consideration.

Following Kitagawa and Ueda \cite{Kitagawa}, we call spin state
$\xi$ to be squeezed iff $V_r(\xi)<s/2$ for some direction $r\perp
\vec{s}$, where
\begin{eqnarray}
\vec{s}=\vec{e}_x \langle S_x \rangle +\vec{e}_y \langle S_y
\rangle +\vec{e}_z \langle S_z \rangle \nonumber
\end{eqnarray}
is the direction of the average spin vector.

This means that in a coordinate system with the $z$-axis along the
average spin vector $\vec{s}$, we always have
\begin{eqnarray}
V_x(\xi)+V_y(\xi)\geq s \label{variance-xy-squeezed}
\end{eqnarray}
in contrast to the spin-coherent state. It is easy to check that
this condition of squeezing (\ref{variance-xy-squeezed}) is valid
for the states (\ref{s=0-state}) and (\ref{s=1-states}), therefore
they are squeezed.

Some spin-squeezed states can be constructed by means of the
squeezing operator \cite{Kitagawa}
\begin{eqnarray}
\mathcal{S}(\xi)= \exp[(\xi^*S_-^2-\xi S_+^2)/2], \nonumber
\end{eqnarray}
so that
\begin{eqnarray}
|\xi \rangle =\mathcal{S}(\xi) ~ |-1 \rangle = -e^{i \varphi} \sin
|\xi| ~ |1 \rangle +\cos |\xi| ~ |-1 \rangle. \nonumber
\end{eqnarray}
Here $\varphi = \arg \xi$. The measure (\ref{measure}) for this
state is
\begin{eqnarray}
\mu (\xi) = |\sin (2|\xi|)|. \nonumber
\end{eqnarray}
Thus, this state is entangled if $|\xi| \neq k \pi/2$, $k=0,1,
\cdots$. At $\xi =\pi /4 +k \pi$, this state coincides with
(\ref{s=1-states}) and hence manifests complete entanglement.

Note that in the case of a single qubit, the squeezing operator is
simply the identity operator. The squeezed states of two qubits
are usually associated with a sort of ``two-mode" squeezing
\cite{Squeezing}.

Thus, for the single spin-qutrit system, coherent states are
unentangled while squeezed states manifest entanglement like in
the case of conventional two-qubit states. Stress that this
correspondence stays within the framework of the definition of
entanglement based on Eqs. (\ref{complete-entanglement}) and
(\ref{entangled-state}).

At the very beginning of the paper, we have stated that the
single-particle entanglement is caused by quantum correlations
between intrinsic degrees of freedom of the particle. The general
picture of those correlations can be revealed through the use of
well known formal correspondence between the states of single
spin-qutrit and two qubits, in other words, of two spin-$1/2$ and
single spin-1. This correspondence is given by the Clebsch-Gordon
decomposition:
\begin{eqnarray}
\mathcal{H}_2 \otimes \mathcal{H}_2=\mathcal{H}_3 \oplus
\mathcal{H}_A, \label{Clebsch}
\end{eqnarray}
Here $\mathcal{H}_2$ denotes the two-dimensional Hilbert space of
states of a single spin-$\frac{1}{2}$, $\mathcal{H}_3$ is the
three-dimensional Hilbert space of spin-1, corresponding to the
symmetric triplet of states in the basis of $\mathcal{H}_2 \otimes
\mathcal{H}_2$, while $\mathcal{H}_A$ corresponds to the
antisymmetric singlet in the basis of $\mathcal{H}_2 \otimes
\mathcal{H}_2$. Denoting the basis in $\mathcal{H}_2$ by
$|\uparrow \rangle $ and $|\downarrow \rangle$, we obtain the
basis in $\mathcal{H}_3$ in the following form
\begin{widetext}
\begin{eqnarray}
|s \rangle = \left\{ \begin{array}{ll} |\uparrow\uparrow \rangle ,
& \mbox{projection of total spin $s=1$} \\ \frac{1}{\sqrt{2}}
(|\uparrow\downarrow \rangle +\downarrow\uparrow \rangle ),&
\mbox{projection of total spin $s=0$} \\ |\downarrow\downarrow
\rangle , & \mbox{projection of total spin $s=-1$} \end{array}
\right. \label{symmetric-states}
\end{eqnarray}
\end{widetext}
while the antisymmetric singlet is
\begin{eqnarray}
|A \rangle = \frac{1}{\sqrt{2}} (|\uparrow\downarrow \rangle
-|\downarrow\uparrow \rangle ). \label{antisymmetric-state}
\end{eqnarray}

If we now assume that  the singlet state
(\ref{antisymmetric-state}) is forbidden because of some physical
reasons, then the system of two qubits becomes exactly equivalent
to the spin-qutrit system. Note, that one of the symmetric states
in (\ref{symmetric-states}) is completely entangled in the
two-qubit sector. This state is clearly equivalent to the state
(\ref{s=0-state}), which is completely entangled in the
spin-qutrit sector as well. On making the further assumption that
spin-qutrit is a local object (particle), we have to associate the
two qubits with intrinsic degrees of freedom of this object.
Simply there is no alternative.

Thus, the single spin-qutrit entanglement can be interpreted in
terms of quantum correlations between the two {\it intrinsic}
qubits under
the following conditions: \\
1. The Hilbert space of two qubits does not contain antisymmetric
states.
\\ 2. System of two qubits is a local one, so that we can neglect
the spatial separation of the qubits and thus interpret them as
intrinsic degrees of freedom of a single ``particle".

In the next Section, we turn to the discussion of possible
physical realizations of spin-qutrit entanglement.

\section{Physical realizations of single spin-qutrit entanglement}

Qubit systems are often associated with two-level atoms, where
quantum correlations between the atoms can be generated either by
photon exchange or by means of dipole-dipole interaction. In the
latter case, decrease of the interatomic distance down to the
lamb-Dicke limit (interatomic separation becomes much shorter than
the wavelength of two-level transition) leads to an effective
discard of the antisymmetric state \cite{Cakir}. Thus, this
two-qubit system behaves like a single spin-qutrit object.

The nice feature of this example is that the reduction of symmetry
\begin{eqnarray}
[\mathrm{SU}(2) \times \mathrm{SU}(2)]_{\mbox{(in 4 dimensions)}}
\rightarrow \mathrm{SU}(2)_{\mbox{(in 3 dimensions)}} \nonumber
\end{eqnarray}
and localization accompany each other.

Another example is provided by the so-called {\it biphoton}, i.e.
by two photons created at once and propagating along the same
direction (see \cite{Biphoton} and references therein). With
respect to polarization, this object represents the
$\mathrm{SU}(2)$ ternary system (spin-qutrit) and is as local as a
single photon. Antisymmetric state with respect to permutations is
forbidden here by the Bosonic nature of photons. Undoubtedly,
biphoton can be split into spatially separated photons, carrying
polarization qubits. But before splitting, it should be considered
as a local spin-qutrit object.

Apologists of the standpoint that entanglement is inherent to
systems with spatially separated parties can say that the above
two examples do not fit the notion of a single particle.
Therefore, we now turn to examples that definitely correspond to a
single particle entanglement.

An important example of the $\mathrm{SU}(2)$ ternary system is
provided by the three-level atom with $\lambda$-type transition
shown in Fig. 1.

\begin{figure}[t]
%\centerline{\scalebox{0.4}{\includegraphics[width=1in]{deneme3_1}}}
\includegraphics[width=8cm]{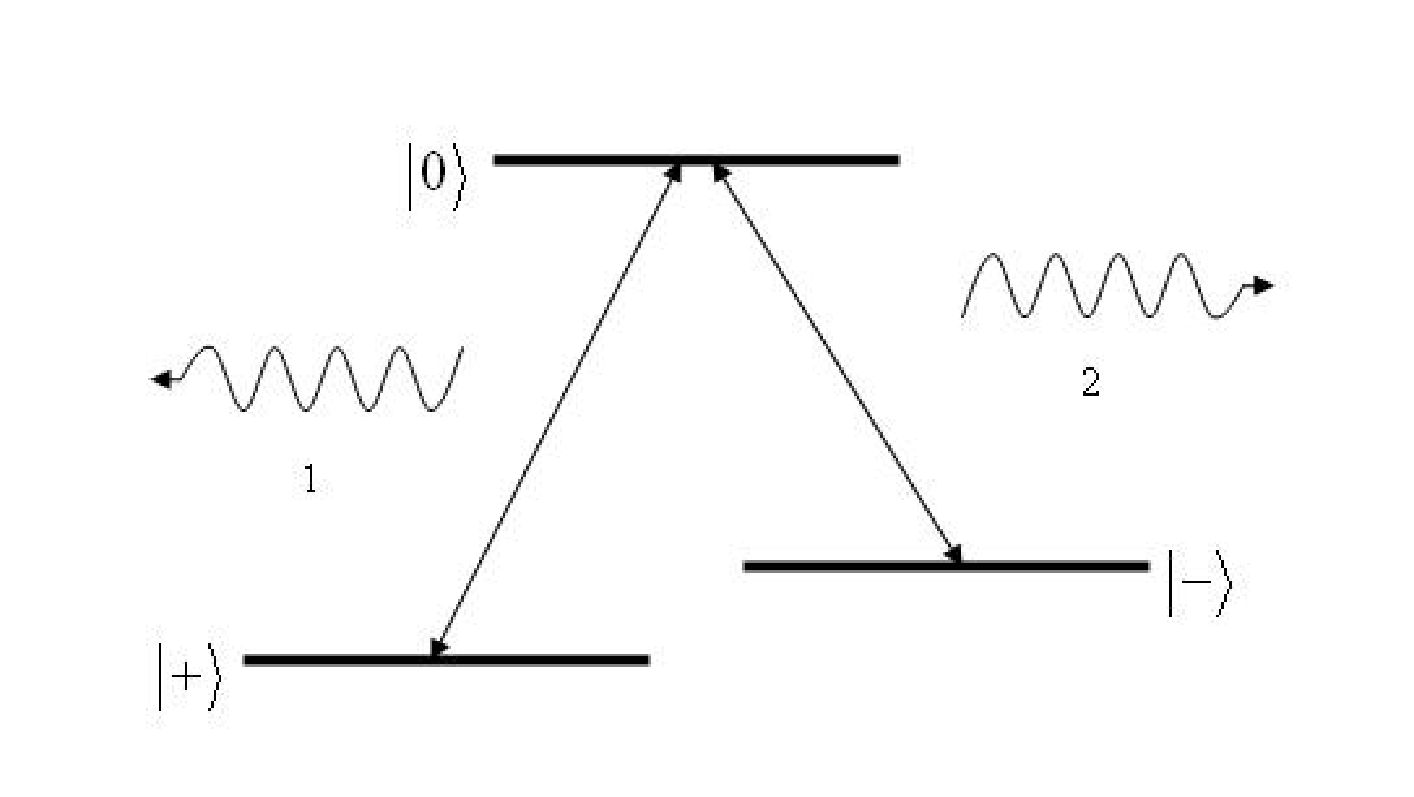}
\caption{\label{fig1}Interaction between $\lambda$-type
three-level atom and two cavity modes.}
\end{figure}
Here the highest excited level can be associated with the state
$|0\rangle$ of ``spin" 1, while the two lower levels with the
states $|+1\rangle$ and $|-1\rangle$, respectively.

The Hamiltonian, describing interaction between the atom and two
cavity modes, has the form
\begin{eqnarray}
H_{int}=g_1R_{0+}a_1+g_2R_{0-}a_2+H.c., \label{Hamiltonian}
\end{eqnarray}
where $g_i$ denotes the corresponding coupling constant,
$R_{bc}=|b\rangle \langle c|$ is the atomic operator, and $a_i$ is
the photon annihilation operator for the field mode $i=1,2$. The
spin operators (\ref{spin-1}) have the form
\begin{eqnarray}
S_x & = & \frac{1}{\sqrt{2}} (R_{+0}+R_{0+}+R_{0-}+R_{-0}),
\nonumber \\ S_y & = & \frac{1-i}{\sqrt{2}}
(R_{+0}+-R_{0+}+R_{0-}-R_{-0}), \label{atom-spin} \\
S_z & = & R_{++}-R_{--} . \nonumber
\end{eqnarray}

In view of the results of Sec. III, the state $|\psi_{in}\rangle =
|0\rangle \otimes |\mathrm{vac}\rangle$ of the atom-field system,
in which the atom is in excited state and cavity field is in the
vacuum state, is completely entangled with respect to the atomic
observables given by Eq. (\ref{atom-spin}). Under influence of the
atom-photon interaction (\ref{Hamiltonian}), this state passes to
the following normalized state
\begin{eqnarray}
\frac{1}{\sqrt{g_1^2+g^2_2}} (g_1|+\rangle \otimes |1_1\rangle
+g_2|-\rangle \otimes |1_2\rangle ) \label{atom-field-ent}
\end{eqnarray}
and vice versa. This state  (\ref{atom-field-ent}) can be
interpreted as the two-qubit state, where one qubit is formed by
the atomic states $|\pm \rangle$ and the second qubit by the
photon states $|1_1\rangle$ and $|1_2 \rangle$. Clearly, this
state is entangled, and the corresponding
concurrence~\cite{Concurrence} has the form
\begin{eqnarray}
\mu = \frac{2|g_1g_2|}{g_1^2+g_2^2} . \nonumber
\end{eqnarray}

This example clearly illustrates decay of the single spin-qutrit
entangled state $|\psi_{in}\rangle$ into the two-qubit
entanglement. The atom and photon qubits can be spatially
separated by cavity leakage.

Another important example of the $\mathrm{SU}(2)$ entanglement of
single particle is provided by the {\it isotriplet} of
$\pi$-mesons. For detailed discussion of this example we refer
recent work~\cite{Kl-Sh-2006}.

The above example of meson isotriplet is similar to the Cooper
pairs in superfluid phases of $^3He$. It is well known that the
atoms of $^3He$ have spin $s=\frac{1}{2}$ each and that the total
spin of a Cooper pair is $s=1$, so that the antisymmetric state of
two atomic qubits is forbidden \cite{Helium}. Note that in the BCS
superconductors where $s=0$, the only allowed pair wave function
is given by the antisymmetric singlet state
(\ref{antisymmetric-state}).

Another simple example of a single particle with spin 1, which can
manifest entanglement, is provided by the deuteron, which is a
nucleus of a deuterium atom, consisting of weakly bounded proton
and neutron \cite{Deuteron}. Note that, unlike $\pi^0$ meson, this
is a stable particle. Each nucleon in the deuteron can be
considered as an intrinsic qubit with respect to its spin
$\frac{1}{2}$. An experimental proof of the existence of
entanglement in deuteron and of the possible use of it for quantum
teleportation of spin states of massive particles has been
reported recently Ref. \cite{Hamieh}.

It is possible to find many other examples, from the spin-1 atoms
like $^{87}\mathrm{Rb}$ and $^{23}\mathrm{Na}$, widely used in
investigation of Bose-Einstein condensation, to the more exotic
systems like vector mesons and three spin-1 gauge bosons in the
standard model \cite{Standard}, in which spin-qutrit entanglement
may be realized.

\section{Conclusion}

We have argued existence of single spin-qutrit entanglement. The
instructive significance of this system is that it allows twofold
consideration as a single spin-1 object and as two qubits, defined
in the symmetric sector of the Hilbert space. This correspondence
allows us to interpret entanglement of single spin-qutrit as
manifestation of quantum correlations between the intrinsic qubit
degrees of freedom. We have shown that entanglement of single
spin-qutrit particle may take place independent of whether or not
the intrinsic qubits can be separated. Thus, the single
spin-qutrit entanglement does not fit conventional requirements of
nonseparability and nonlocality. At the same time, the single
spin-qutrit entanglement has all physical features of two-qubit
entanglement. In particular, entangled states of a single
spin-qutrit are squeezed  and unentangled states are coherent like
in the case of bipartite systems.

We have discussed a number of physical objects that can be
prepared in entangled spin-qutrit states. We have shown that the
physical condition of complete entanglement as extreme of quantum
fluctuations can be important for understanding of low stability
of entangled states of particles.

The obtained result about the single-particle entanglement for the
spin-qubit system is clearly valid for all systems with high
enough dynamic symmetry $\mathrm{SU}(N)$ at $N\geq 3$.

The obtained results show distinctly that the physical definition
of entanglement \cite{Kl-Sh-2006} based on definition of basic
observables and their quantum fluctuations, is more general than
the previous definitions that appeal to nonlocality and
nonseparability. As we have shown, using true and spin qutrits as
an illustrative example, the presetting of basic observables plays
crucial role in the description of entanglement. In particular, it
defines specific relativity of entanglement with respect to
dynamic symmetry of physical system. The definition in terms of
the variational principle (\ref{complete-entanglement}) can be
used for investigation of entanglement of different physical
objects, including elementary particle, quasi-particle excitations
in condensed matter and so on. Thus, it essentially broaden the
applicability of this notion beyond the bounds of quantum
information. It is possible to say that the association of
entanglement with quantum uncertainties of basic observables makes
this notion to be ubiquitous in physics.

The possibility of experimental observation of single-particle
entangled states represents a problem of high importance and
deserves special discussion. Let us only note that the decay of a
single entangled $\mathrm{SU}(2)$ qutrit into two entangled qubits
may be used for this aim.

In our analysis, we have used a general approach to quantum
entanglement \cite{Kl-Sh-2006}, which assigns the primary
importance to the dynamic symmetry properties of physical systems.

We have restricted our consideration by pure states. In the
future, we hope to extend our approach on the mixed states. In
particular case of single $\mathrm{SU}(2)$ qutrit, the mixed state
entanglement can be quantified in the same way as the two-qubit
entanglement (see Appendix C). So far, the measure of mixed-state
entanglement is known only for two qubits (see the first reference
in~\cite{Concurrence}). The principle difficulty here is that the
total variance of mixed states contains contributions of both
quantum and classical (statistical) uncertainties. The problem of
detachment of the two principally different contributions deserves
special discussion. The ideas related to the Wigner-Yanase quantum
information \cite{Wigner-Yanase,Wigner} may be useful here.

\section*{Acknowledgements}

Part of this work was carried out (by M.A.C.) at the Jet
Propulsion Laboratory, California Institute of Technology, under a
contract with the National Aeronautics and Space Administration
(NASA). M.A.C. also acknowledges support from the Oak Ridge
Associated Universities (ORAU) and NASA.

\section*{Appendix A: Classical realism}

The problem of classical realism can be formulated in the
following way. It is known that the measurement of a quantum
mechanical observable $X_i$ in a given state $\psi$ produces
random quantity $x_i$ determined by expectations $\langle
\psi|f(X_i)|\psi \rangle$ of all functions $f(x_i)$ (it is usually
enough to consider only moments $f(x_i)=x_i^n$ (see Ref.
\cite{Neumann}). In the case of commuting quantum observables
$[X_i,X_j]=0$, the corresponding random quantities $x_i$ and $x_j$
have the same joint probability distribution. Einstein's idea of
classical realism \cite{EPR} assumes that all quantum observables
have the same {\it hidden} joint distribution independent of
whether they are commuting or not.

Bell's approach to prove nonexistence of hidden variables is based
on formulation of certain ``classical" conditions on measurement
of quantum observables and check of their violation in quantum
mechanics \cite{Bell}. Note that, from the mathematical point of
view, problem of Bell's conditions lies within the problem of
distributions with given margins
\cite{Klyachko-2002,Klyachko-2006} and that practically all known
and many still unknown Bell-type inequalities were obtained in
mathematics even before the formulation of problem by Bell.

It has been proved in Ref. \cite{Klyachko-2002} that an
irreducible quantum system with dynamic symmetry group $G$ of rank
$\mathrm{rk}(G) \geq 2$ is incompatible with classical realism,
i.e. violates Bell-type conditions. Since
$\mathrm{rk}(\mathrm{SU}(n))=n-1$, state of a single true qutrit
with the dynamic symmetry group $G=\mathrm{SU}(3)$ violate Bell's
conditions of classical realism. It has been shown in Section III
that the true qutrit states do not manifest entanglement.

\section*{Appendix B: Three-qubit entanglement}

Three-qubit states may manifest entanglement of two different
types. Namely, entanglement caused by correlations of all three
qubits and entanglement due to correlation between pair of parts
\cite{Miyake}. The three-qubit entanglement is measured by means
of 3-tangle \cite{Miyake,Wootters-1}, which for the general state
\begin{eqnarray}
|\psi \rangle = \sum_{k,\ell,m =0}^1 \psi_{k\ell m} |k \ell m
\rangle, \quad \sum_{k, \ell ,m=0}^1 |\psi_{k \ell m}|^2=1,
\nonumber
\end{eqnarray}
has the form
\begin{widetext}
\begin{eqnarray}
\tau
(\psi)&=&4|\psi^2_{000}\psi^2_{111}+\psi^2_{001}\psi^2_{110}+\psi^2_{010}\psi^2_{101}
+\psi^2_{100}\psi^2_{011}-2(\psi_{000}\psi_{001}\psi_{110}\psi_{111}+\psi_{000}\psi_{010}\psi_{101}\psi_{111}
+\psi_{000}\psi{100}\psi_{011}\psi_{111} \nonumber
\\ &+&\psi_{001}\psi_{010}\psi_{101}\psi_{110}
+\psi_{001}\psi_{100}\psi_{011}\psi_{110}
+\psi_{010}\psi_{100}\psi_{011}\psi_{101})+4(\psi_{000}\psi_{011}\psi_{101}\psi_{110}
+\psi_{001}\psi_{010}\psi_{100}\psi_{111})| \label{3-tangle}
\end{eqnarray}
\end{widetext}

According to classification by Miyake \cite{Miyake}, the following
three states
\begin{eqnarray}
|GHZ \rangle & = & \frac{1}{\sqrt{2}} (|000 \rangle +|111 \rangle
), \label{GHZ} \\ |W \rangle & = & \frac{1}{\sqrt{3}} (|011
\rangle +|101 \rangle +|110 \rangle ), \label{W} \\ |Bi \rangle &
= & \frac{1}{\sqrt{2}} (|011 \rangle +|101 \rangle)  \label{Bi}
\end{eqnarray}
are generic for the three SLOCC-nonequivalent classes in the
eight-dimensional Hilbert space. The {\it non-separable} states
from the GHZ class manifest only three-partite entanglement
(3-tangle (\ref{3-tangle}) has nonzero values for the states from
this class), while any pair of qubits is unentangled. The latter
can be checked by reduction of the three-qubit density matrix
$\rho_{GHZ}=|GHZ \rangle \langle GHZ|$ to the two-qubit mixed
state ${\rho'}_{GHZ}=\mathrm{Tr}_{single} \rho_{GHZ}$, where
$\mathrm{Tr}_{single}$ denotes trace over one of the parts, with
the subsequent calculation of the concurrence, which in this case
always have zero value.

In turn, the {\it non-separable} states from the W class always
have zero 3-tangle and hence do not manifest three-partite
entanglement. In turn, any bipartite reduced state with the
density matrix ${\rho'}_W=\mathrm{Tr}(|W\rangle \langle W|)$ has
nonzero concurrence and therefore shows bipartite entanglement.

Finally, the {\it separable} states from the Bi class are similar,
in a sense, to the W states. Namely, they always have zero
3-tangle while manifest bipartite entanglement (for two given
qubits only).

Thus, the nonseparability (separability) of the three-qubit states
does not indicate identically the presence (absence) of
entanglement and its type in contrast to the bipartite systems.

In the latter case, entanglement is usually associated with
specific behavior of entropy of the reduced state single-part
state. Namely, entanglement of two qubits with statistical state
$\rho_{AB}$ exists if $H(\rho_A)=H(\rho_B) \neq 0$, where
$H(\rho)=-\mathrm{Tr}(\rho \log \rho)$ is the von Neumann entropy.
The maximal (complete) entanglement corresponds to the reduced
states that have the diagonal form
\begin{eqnarray}
\rho_r= \left(\begin{array}{cc}1/2 & 0 \\ 0 & 1/2 \end{array}
\right), \quad r=A,B, \nonumber
\end{eqnarray}
in a certain basis of the two-dimensional single-qubits space
\cite{horodecki}.

In the above case of three-qubits, these conditions do not work.
Consider as an example the GHZ-type state of the form $|\Psi
\rangle = x|000 \rangle +y|111 \rangle$, $x^2+y^2=1$. It is seen
that 3-tangle (\ref{3-tangle}) $\tau (\Psi)=4x^2y^2=4x^2(1-x^2)$,
so that the state is entangled (in the three-part sector) for all
$x \in (0,1)$. The reduced two-qubit density matrix for any pair
of qubits has the form
\begin{eqnarray}
\rho_{R}=x^2|00 \rangle \langle 00|+(1-x^2)|11 \rangle\langle 11|
\nonumber
\end{eqnarray}
with the corresponding von Neumann entropy
\begin{eqnarray}
H(\rho_R)=-x^2 \log x^2-(1-x^2 \log (1-x^2). \nonumber
\end{eqnarray}
Subsequent reduction of $\rho_R$ to the single-qubit state
\begin{eqnarray}
\rho_{RR}=x^2|0 \rangle \langle 0|+(1-x^2)|1 \rangle \langle 1|
\nonumber
\end{eqnarray}
obviously leads to the same von Neumann entropy $H(\rho_R)$ as
$\rho_R$, although there is no two-qubit entanglement in the
state. Similar behavior, showing unfitness of the reduced state
entropy as a general measure of entanglement, is manifested by the
$W$ and Bi states of three qubits as well.

\section*{Appendix C: Mixed state entanglement}

Amount of entanglement carried by a mixed single spin qutrit state
can be calculated in the same way as for two qubits through the
use of Wootters'  concurrence \cite{Concurrence}. Namely
\begin{eqnarray}
\mu(\rho)=\max(0,\lambda_1-\lambda_2-\lambda_3),
\label{concurrence-mixed}
\end{eqnarray}
where $\lambda_i$ are the square roots of the eigenvalues of the
$(3 \times 3)$ matrix $\rho F \rho^* F$, in decreasing order. Here
the ``spin-flip" transformation matrix $F$ is defined for the spin
qutrit state as follows
\begin{eqnarray}
F= \left( \begin{array}{lll} 0 & 0 & -1 \\ 0 & 1 & 0 \\ -1 & 0 & 0
\end{array} \right) \nonumber
\end{eqnarray}
It corresponds to the Wootters' spin-flip transformation $\sigma_y
\otimes \sigma_y$ defined in the symmetric sector of the
four-dimensional Hilbert space.

For example, the concurrence (\ref{concurrence-mixed}) of the
``symmetric" Werner state
\begin{eqnarray}
\rho_W= \frac{x}{3} (|+1 \rangle \langle +1|+|0 \rangle \langle
0|+|-1 \rangle \langle -1|) +(1-x)|0 \rangle \langle 0|, \nonumber
\end{eqnarray}
which represents superposition of completely mixed and completely
entangled states, has the form
\begin{eqnarray}
C(\rho_W)= \left\{ \begin{array}{ll} (1-4x/3), & \mbox{at $0 \leq
x <3/4$} \\ 0, & \mbox{at $3/4 \leq x \leq 1$} \end{array} \right.
\nonumber
\end{eqnarray}
Remind that in the case of conventional two-qubit Werner state
\cite{Werner} the concurrence has the form
\begin{eqnarray}
C(\rho_{W_{4d}})= \left\{ \begin{array}{ll} (1-3x/2), & \mbox{at
$0 \leq x <2/3$} \\ 0, & \mbox{at $2/3 \leq x \leq 1$} \end{array}
\right. \nonumber
\end{eqnarray}
Thus, entanglement of ``symmetric" Werner state survives at higher
admixture of completely chaotic state than that of Werner state in
the whole space $\mathcal{H}_2 \otimes \mathcal{H}_2$.

\section*{Appendix D: Violation of Bell-type condition by single spin $1$}

Spin 1 state space can be represented with complexification of
Euclidian space of $\mathcal{H}=\mathbb{R}^3\otimes\mathbb{C}$
where $SU(2)\sim SO(3)$ acts on $\mathcal{H}$ by rotations
$\mathbb{R}^3$. It inherited from $\mathbb{R}^3$, the bilinear
scalar and cross products are denoted by $(x,y)$ and $[x,y]$
respectively. Spin projection onto direction $\ell\in
\mathbb{R}^3$ becomes

$$S_\ell\psi=i[\ell,\psi].$$

Spin projection has one real eigenstate $|0\rangle=\ell$ with
eigenvalue $0$ and two imaginary eigenstates
$|\pm1\rangle=\frac{1}{\sqrt{2}}(m\pm in)$ with eigenvalues
$\pm1$, where $(\ell,m,n)$ is orthonormal basis of $\mathbb{R}^3$
(e.g. coordinate vectors $i,j,k$).

Real vector $|0\rangle$ is completely entangled spin state, while
complex ones are coherent, as we discussed in the paper.

%By a rotation every spin 1 state can be put into the {\it
%canonical form}

%\begin{equation}
%\label{canonic}
%\psi=m\cos\varphi+in\sin\varphi,\quad0\le\varphi\le \frac{\pi}{4}.
%\end{equation}

Let us define operator $R_{\ell}=2S_\ell^2-1$. While $S_\ell$ has
three eigenvalues, $R_\ell$ has only two eigenvalues $\pm1$.
Observe that $R_\ell^2=1$ and operators $R_\ell$ and $R_m$ commute
iff $\ell\perp m$. Then they have joint probability distribution,
and are simultaneously measurable observables.

Consider a cyclic quintuplet of unit vectors
$\ell_i\in\mathbb{R}^3$, $i \mathrm{\;mod\;} 5$,
 such that $\ell_i\perp\ell_{i+1}$, and call it {\it pentagram\/}.

\begin{figure}[t]
%\centerline{\scalebox{0.3}{\includegraphics{pent4.eps}}}
\includegraphics[width=8cm]{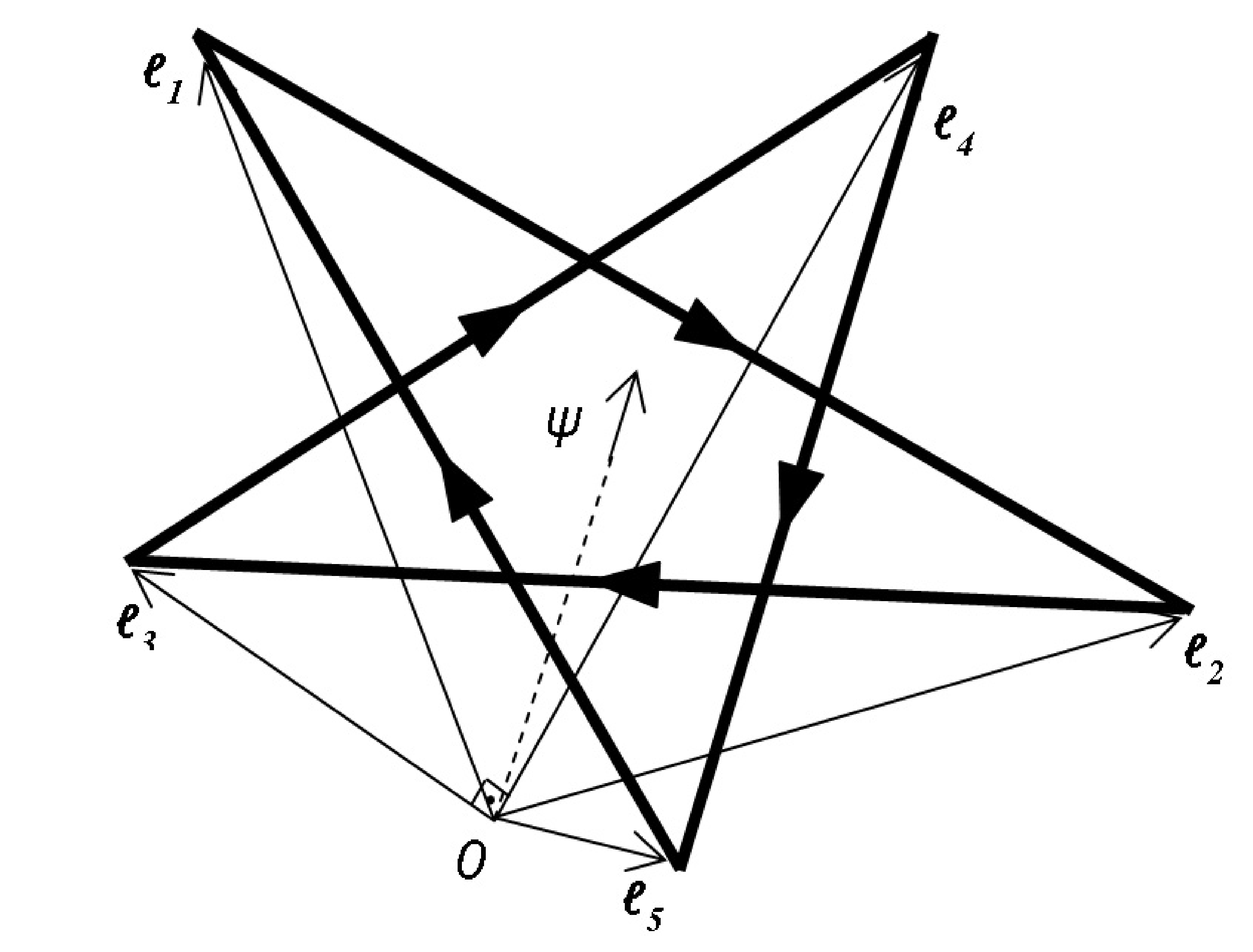}
\caption{\label{fig2}Regular planar pentagram.}
\end{figure}

Put $R_i:=R_{\ell_i}$. Then $[R_i,R_{i+1}]=0$ and for all possible
eigenvalues $r_i=\pm1$ of observable  $R_i$

$$(r_1r_2)(r_2r_3)(r_3r_4)(r_4r_5)(r_5r_1)=1.$$
Among those five monomials at least one is positive. So the
following inequality holds
\begin{equation}\label{spin_bell}
r_1r_2+r_2r_3+r_3r_4+r_4r_5+r_5r_1+3\ge0.
\end{equation}

Iff all $R_i$ would have a joint distribution then taking average
of (\ref{spin_bell}) one get Bell's type inequality
\begin{widetext}
\begin{equation}\label{pent_bell}
\langle\psi|R_1R_2|\psi\rangle+\langle\psi|R_2R_3|\psi\rangle+
\langle\psi|R_3R_4|\psi\rangle+ \langle\psi|R_4R_5|\psi\rangle+
\langle\psi|R_5R_1|\psi\rangle+3\ge0
\end{equation}
\end{widetext}
for testing classical realism.

Using the identity $R_i=1-2|\ell_i\rangle\langle\ell_i|$ one can
recast it into geometrical form
\begin{equation}\label{spin_bell1}
\sum_{i\mathrm{\;mod\;}5}|\langle\ell_i,\psi\rangle|^2\le
2\Longleftrightarrow\sum_{i\mathrm{\;mod\;} 5}\cos^2\alpha_i\le
2,\qquad\alpha_i=\widehat{\ell_i\psi}.
\end{equation}

Completely entangled spin states easily violate this inequality.
Say for regular pentagram and $\psi\in\mathbb{R}^3$ directed along
its axis of symmetry one gets
$$\sum_{i\mathrm{\;mod\;}5}\cos^2\alpha_i=\frac{5\cos\pi/5}{1+\cos\pi/5}\approx2.236>2.$$

It turns out that every non-coherent spin state violates
inequality (\ref{spin_bell1}) for an appropriate pentagram, and
the coherent states pass this test for any pentagram, see the Ref.
\cite{Klyachko-2007} for details.

\end{document}